\documentclass[fleqn,twoside]{article}
\usepackage{espcrc2}

\usepackage{graphicx}
\usepackage[figuresright]{rotating}

\newcommand{\AmS}{{\protect\the\textfont2
  A\kern-.1667em\lower.5ex\hbox{M}\kern-.125emS}}

\title{Top Quark Mass Measurements at CDF}
\author{Pedro A. Movilla Fernandez\address[LBNL]{Lawrence Berkeley National Laboratory, 
        1 Cyclotron Road, Berkeley, California 94720, U.S.A.}  ({\em on behalf of the CDF Collaboration})}

\newcommand{\eg}   {{\it e.g.}}

\newcommand{\vs}   {{\it vs.}}
\newcommand{\etal}   {{\it et al.}}

\newcommand{\MET} {\ensuremath{{\not}{E_\mathrm{T}}}}
\newcommand{\rs}   {\ensuremath{\sqrt{s}}}
\newcommand{\mH}    {\ensuremath{m_H}}
\newcommand{\mW}    {\ensuremath{m_{W}}}
\newcommand{\mt}    {\ensuremath{m_{t}}}
\newcommand{\tbbar} {\ensuremath{{t\bar{b}}}}
\newcommand{\ttbar} {\ensuremath{{t\bar{t}}}}
\newcommand{\ppbar}              {\ensuremath{p\overline{p}}}
\newcommand{\bbbar}     {\ensuremath{{b\bar{b}}}}
\newcommand{\ccbar}     {\ensuremath{{c\bar{c}}}}
\newcommand{\qqbar}     {\ensuremath{{q\bar{q}}}}
\newcommand{\qqbarprime}     {\ensuremath{{q\bar{q}^\prime}}}

\newcommand{\pt}               {\ensuremath{p_\mathrm{t}}}

\newcommand{\ET}               {\ensuremath{E_\mathrm{T}}}

\newcommand{\gevcc}     {\ensuremath{\mathrm{GeV}/c^2}}

\newcommand{\invpb}     {\ensuremath{\mathrm{pb}^{-1}}}
\newcommand{\invfb}     {\ensuremath{\mathrm{fb}^{-1}}}

\newcommand{\invcms}     {\ensuremath{\mathrm{cm}^{-2}\mathrm{s}^{-1}}}

\newcommand{\nat}[1]{Nature {\bf #1}}
\newcommand{\prl}[1]{Phys. Rev. Lett. {\bf #1}}

\newcommand{\prd}[1]{Phys. Rev. D {\bf #1}}

\newcommand{\plet}[1]{Phys. Lett. {\bf #1}}

\begin{document}

%%%%%%%%%%%%%%%%
%%% Abstract %%%
%%%%%%%%%%%%%%%%

\begin{abstract}
Recent measurements of the mass of the top quark ($t$) are presented
using 162\,\invpb\ of data of \ppbar\ collisions at $\rs\,=1.96\,$TeV
collected by the CDF detector at the Tevatron collider during Run~II.
The analyses focus on the semi-leptonic decay mode $\ttbar\to (bl\nu)
(\overline{b}\qqbar)$ with one or two identified bottom quarks ($b$).
The Template Method reconstructs the invariant mass of the
top quark in each event.  The Multivariate Template Method
enhances this approach by adding information on the event
topology. The Dynamical Likelihood Method discriminates between
possible mass values using top quark decay observables and attempts to
use the maximum amount of information on top quarks provided by the
Standard Model.  All three methods produce similar results. The
Dynamical Likelihood Method yields a top quark mass of
$177.8^{+4.5}_{-5.0}\,\mathrm{(stat)}\pm6.2\,\mathrm{(sys)}\,\gevcc$.
\vspace{7pt}
\end{abstract}

\maketitle

%%%%%%%%%%%%%%%%%%%%
%%% Introduction %%%
%%%%%%%%%%%%%%%%%%%%

\section{INTRODUCTION}

The first experimental observation of the top quark ($t$) by the CDF
and D\O\ collaborations at the Tevatron about ten years
ago~\cite{bib:topdiscovery} represented an impressive confirmation of
the Standard Model (SM) and marked the beginning of a successful top
quark physics program at the Tevatron.  Properties of the top quark
were first investigated using 110\,\invpb of data at $\rs=1.8\,$TeV
collected in the Run~I period between 1992 and 1996.  The top quark
mass, \mt, was measured in all decay topologies (\eg\
\cite{bib:cdfrun1,bib:d0run1}) arising from $\ppbar\to\ttbar X$ final states.
After a recent reanalysis of D\O\ Run~I data~\cite{bib:d0run1new}, the
collaborations updated the combined Run~I value to
$\mt=178.0\pm4.3\,\gevcc$~\cite{bib:ewg}.

The precise knowledge of \mt\ is of great interest for particle
physics. It is a fundamental parameter of the SM, but more
importantly, the extremely high mass gives the top quark particular
relevance in the calculation of other SM parameters. Electroweak
corrections to the $W$ propagator introduce a quadratic dependence on
the $W$ boson mass \mW\ and a logarithmic dependence on the mass of
the long-hypothesized but still unobserved Higgs boson, \mH. With the
current precision level of 2.5\% for \mt\ and $<0.1\%$ for \mW, the
most likely value for \mH\ obtained from global SM fits is
$\mH=114^{+69}_{-45}\,\gevcc$, and the upper 95\% CL limit is
$260\,\gevcc$~\cite{bib:ewg}. A more accurate determination of
\mt\ would provide the opportunity for stringent consistency tests of
the SM.

The precision measurement of the top quark mass is therefore a very
active topic in the current physics program of CDF in Run~II.  The
upgraded Tevatron complex started in 2001, producing \ppbar\
collisions at \rs=1.96\,TeV with steadily increasing instantaneous
luminosities up to a record of $\sim10^{32}\,\invcms$ (as of the time
of this report).  The higher collision energy led to an increase of
the SM production \ttbar\ cross-section of~35\%~\cite{bib:xsec}.  The
CDF detector~\cite{bib:cdf2det} underwent substantial improvements in
the geometrical acceptance of the tracking and calorimeter systems as
well as of the silicon vertex detector which is crucial for the
identification of $B$-hadrons from $t$ decays.

All \ttbar\ decay modes are being investigated intensively. Here we
report on three recent CDF measurements of \mt\ in the lepton plus
jets channel using 162\,\invpb data collected between March 2002 and
September 2003.

%%%%%%%%%%%%%%%%%%%%%%%%
%%%  Event Selection %%%
%%%%%%%%%%%%%%%%%%%%%%%%

\section{EVENT SELECTION}

According to the Standard Model, top quarks at the Tevatron are mainly
produced in pairs through quark-anti-quark annihilation (85\%) or
gluon-gluon fusion (15\%), followed by a prompt decay $t\to Wb$ with a
branching ratio BR$\sim100\%$. The subsequent $W$ decays define the
event signature.  In the decay mode studied here, one $W$ decays into
two quarks, and the other $W$ decays into an electron ($e$) or a muon
($\mu$) and the corresponding neutrino ($\nu$). This channel has a
manageable signal to background ratio $S/B\sim\mathcal{O}(1)$ and a
satisfactory $BR\sim\frac{8}{27}$ that is viewed as a good compromise
between all available decay modes.\footnote{The {\em dilepton mode}
(with $e$ or $\mu$) has $S/B\sim10$, $BR\sim\frac{4}{81}$, the {\em
all-hadronic} mode has $S/B<\frac{1}{10}$, $BR\sim\frac{4}{9}$.}

The final state has a spherical topology due to \ttbar\ production
near threshold and is characterized by four jets, one charged lepton
with high transverse energy (\ET), and a sizable amount of missing
transverse energy (\MET) due to the undetectable $\nu$. Top quark
candidate events are selected by requiring exactly one isolated $e$ or
$\mu$ with $\ET>20$\,GeV and $\MET>20$\,GeV. Furthermore, at least
three jets with $\ET>15$\,GeV and one jet with $\ET>8$\,GeV have to be
present.  The DLM analysis (explained below) requires exactly four
jets with $\ET>15$\,GeV. The jets must lie in the pseudo-rapidity
region $|\eta|<2$ so that they are well measured by the detector.
Finally, one or two displaced secondary vertices (originating from the
decay of long-lived $B$-hadrons) are identified by the silicon vertex
detector (``$b$-tags''). This requirement improves the sample purity
by a factor of 3-4 compared to the non-$b$-tagged case.

Dominant background sources are $W\bbbar$ and $W\ccbar$ final states,
non-$W$ QCD processes and events where one jet is misidentified as a
$b$-jet.  Minor sources of background are $WW$/$WZ$, $Wc$ and single
top quark (\tbbar) events.

%%%%%%%%%%%%%%%%%%%%%%%%%%%%%%
%%%  Top Mass Measurements %%%
%%%%%%%%%%%%%%%%%%%%%%%%%%%%%%

\section{TOP MASS MEASUREMENTS}

\subsection{Template Method}
The Template Method (TM) closely follows the traditional mass measurement strategy
in Run~I.  To get the momenta of the daughter particles of the assumed
initial \ttbar\ configuration, the raw jet energies are corrected for
detector effects and artifacts due to the limited jet cone size, as
well as for hadronization and for various non-\ttbar\ contributions of
the same event.  One part of the corrections are directly derived from
the data, another part is derived using a Herwig \ttbar\ Monte
Carlo~\cite{bib:herwig} and a simulation of the CDF detector.  The
analysis treats the four leading \ET\ jets as quark jets from the
\ttbar\ system. The event reconstruction is complicated by the
combinatorial problem of correctly assigning these jets to the four
initial partons, yielding six (two) solutions (``permutations'') for
the single (double) $b$-tag case. The number of solutions doubles
because of the twofold ambiguity of the longitudinal component of the
$\nu$ momentum.

\begin{figure}[t]
\vspace{-18pt}
\includegraphics[width=.42\textwidth]{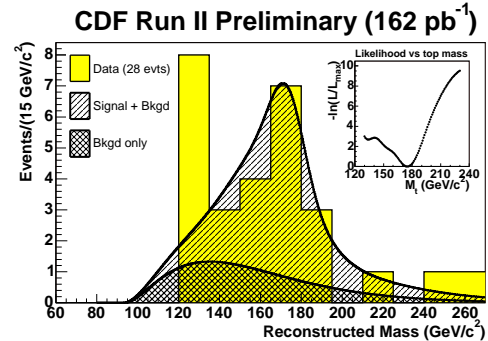}
\vspace*{-32pt}
\caption{\small Reconstructed top quark mass data 
compared with the combined signal plus background fit.
\vspace*{-25pt}}
\label{fig:tm}
\end{figure}

For each configuration, an invariant top quark mass is reconstructed
by minimizing a $\chi^2$ expression calculated using the transverse
momenta of the jets, the charged lepton, and the missing energy. The
$\chi^2$ takes into account the detector resolution as well as the decay widths of
the $t$ and the $W$,  and also incorporates various
kinematic constraints (\eg\ $m_t=m_{\bar{t}}$, $M_{W^+}=M_{W^-}$). The
solution with the smallest $\chi^2$ value 
%($\chi^2_0$) 
is taken as the ``reconstructed top mass'' of the candidate
event. This quantity is used to build templates for signal Monte Carlo
samples with varying generated \mt\, but also for various background
samples.  The signal templates are parametrized by continuous
functions of $\mt$.  The templates serve as probability densities to
be compared with the data in order to derive a top quark mass
probability for each event. The most likely value for \mt\ is
extracted using an unbinned likelihood fit that maximizes the joint
probability density (= product of all event-by-event probability
densities) for the whole sample of reconstructed masses. It also
includes a background fraction constraint.

With a sample of 28 candidate events (Fig.~\ref{fig:tm}) and
$6.8\pm1.2$ expected background events~\cite{bib:petra_merkel} the
method yields
$\mt=174.9^{+7.1}_{-7.7}\mathrm{(stat)}\pm6.5\mathrm{(sys)}\,\gevcc$.
The systematic error is dominated by uncertainties of the jet energy
scale ($\pm6.3\,\gevcc$) which mainly arises from the calibration of
the CDF calorimeter to a uniform response and setting the absolute jet
energy scale.

%%%%%%%%%%%%%%%%%%%%%%%%%%%%%%%%%%%%
%%% Multivariate Template Method %%%
%%%%%%%%%%%%%%%%%%%%%%%%%%%%%%%%%%%%

\subsection{Multivariate~Template~Method}

The Multivariate~Template~Method (MTM) has three main features: 1) The jet energy scale (JES) is adjusted
event-by-event.  2) Three types of signal templates are used to handle
the combinatorial problem. 3) Other kinematic variables, in addition
to \mt, are used to improve S/B discrimination. The method performs
the same generic jet corrections as TM but uses a refined kinematic
fitter for the event reconstruction. In particular it includes a JES
factor calibrated by the reconstruction of the $W\to\qqbarprime$ decay
using a $W$ mass constraint. On average this partially compensates for
systematic JES shifts but also increases the statistical error because
of event-by-event fluctuations of the JES. The fitter contains a
JES constraint which allows the balance of JES systematics \vs\
statistical uncertainties.  Further kinematic constraints similar to
TM are imposed.  Only the neutrino solution for which the $t$ and
$\bar{t}$ masses are closest is taken.

Three signal templates are used to handle jet-parton combinatorics:
correct permutation samples, incorrect permutation samples, and
incorrect jet assignment samples (where \eg\ a gluon jet is assigned to a
quark). It can be shown that the knowledge of the event template type
improves the mass resolution by a factor of $\sim1.7$ in an ideal
scenario with no background.  MTM attempts to predict the correct
permutation probability by using the differences between the smallest
$\chi^2$ and the $\chi^2$ values of the other permutations, and by
using kinematic information with weak \mt\ dependence, specifically
the angle between the lepton and $b$-quark in the $W\to l\nu$ rest
frame, and \ttbar\ spin correlation variables. These probabilities are
used to weight the signal templates accordingly.

In addition to the reconstructed top quark mass, further kinematic
variables with high S/B separation power are used to construct
multidimensional templates. From studies of various histogram
divergence measures it turns out that the scalar sum of the transverse
momenta of the four leading jets ($P_\mathrm{T4}$) is a promising
variable to distinguish signal from noise. The templates are generated
using Kernel Density Estimation~\cite{bib:kde}, a non-parametric
density reconstruction technique particularly suited for multivariate
densities.

\begin{figure}[t]
\vspace{-11pt}
\includegraphics[width=.44\textwidth]{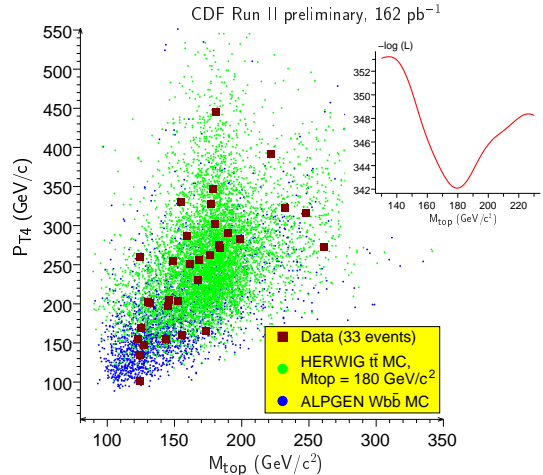}
\vspace*{-31pt}
\caption{\small Reconstructed mass \vs\ $P_\mathrm{T4}$ for the data
and for signal and background expectation.
\vspace*{-14pt}}
\label{fig:mtm}
\end{figure}

The top quark mass likelihood is expressed in terms of multivariate
probability densities for signal and background and considers the
three types of signal templates.  The background fraction is allowed
to float freely.  The JES constraint is optimized w.r.t. the expected
total error.  A maximum likelihood method applied to a sample of 33
selected events (Fig.~\ref{fig:mtm}) yields
$\mt=179.6^{+6.4}_{-6.3}\mathrm{(stat)}\pm6.8\mathrm{(sys)}\,\gevcc$
and a most probable background fraction of $0.34\pm0.14$.

%%%%%%%%%%%%%%%%%%%%%%%%%%%%%%%%%%%
%%% Dynamical Likelihood Method %%%
%%%%%%%%%%%%%%%%%%%%%%%%%%%%%%%%%%%

\subsection{Dynamical~Likelihood~Method}

The Dynamical~Likelihood~Method (DLM) is an original CDF method~\cite{bib:DLM} that attempts to use the
full amount of information about the \ttbar\ process provided by the
Standard Model. The top quark mass likelihood $L^{(i)}(m_t)$ for the
$i^{th}$ event is given by
\begin{eqnarray*}
L^{(i)}(m_t)
 &=&  
  \int_{}^{} {\sum_\mathrm{perm}}\;{\sum_\mathrm{\nu sol.} }
  \frac{\scriptstyle 2\pi^4}{\scriptstyle \mathrm{flux}}|
  \mathcal{M}|^2F(z_1,z_2)f(\pt)\\
 & &
 \times w(\mathbf{x},\mathbf{y};m_t) \;\mathrm{d}\mathbf{x}
\end{eqnarray*}
where $\mathbf{x}$ denotes the parton momenta and $\mathbf{y}$ denotes
jet observables. $F(z_1,z_2)$ is the parton distribution function for
the incoming partons, and $f(\pt)$ is the probability that the \ttbar\
system has received a transverse momentum \pt\ due to initial state
radiation.  The matrix element $\mathcal{M}$ for the \ttbar\
production and decay process is linked to the four-jet configuration
using transfer functions $w(\mathbf{x},\mathbf{y};m_t)$ (TF). These are
Bayesian probabilities that a \ttbar\ parton configuration
$\mathbf{x}$ with a given \mt\ was generated when $\mathbf{y}$ was
reconstructed.  The TFs are functions of
$(\ET^{\mathrm{parton}}-\ET^{\mathrm{jet}})/\ET^{\mathrm{parton}}$ and
parametrized in jet \ET\ and $\eta$ bins. They are calculated for $W$
or $b$ jets separately using a Herwig \ttbar\ MC for a fixed top quark
mass.  DLM performs some of the jet corrections as TM and MTM (\eg\
relative calorimeter response, calorimeter to particle correction),
while it handles the transition from partons to particles via TFs.

\begin{figure}[t]
\vspace{-14pt}
\centerline{\includegraphics[width=.43\textwidth]{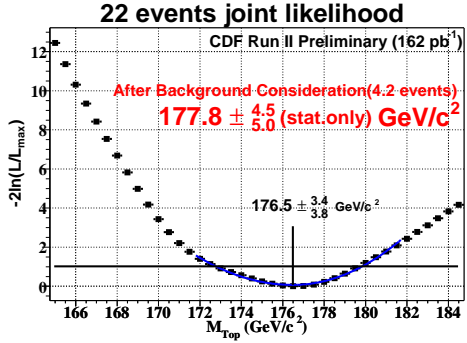}}
\vspace{-28pt}
\caption{\small DLM joint negative log likelihood.
\vspace{-20pt}
}
\label{fig:dlm}
\end{figure}

The likelihood function $L^{(i)}$ sums {\it all} permutations and
neutrino solutions.  The most likely top quark mass is obtained by
maximizing $\Pi_iL^{(i)}(\mt)$. The joint negative log likelihood
curve extracted from a sample of 22 selected events is shown in
Fig.~\ref{fig:dlm}.  It should be noted that the likelihood considers
only the signal but not the background matrix elements. The background
contribution is hence minimized by requiring exactly four
jets. Residual background effects, but also the top quark mass
dependence of the TFs, are taken into account by applying an MC-based
correction (``mapping'' function) to the maximum likelihood mass.
Using $4.2\pm0.2$ expected background events, this method yields
$\mt=177.8^{+4.5}_{-5.0}\mathrm{(stat)}\pm6.2\mathrm{(sys)}\,\gevcc$.

%%%%%%%%%%%%%%%%%%%%%%%%%%%
%%% Summary and Outlook %%%
%%%%%%%%%%%%%%%%%%%%%%%%%%%

\section{SUMMARY AND OUTLOOK}

Three recent CDF measurements of the top quark mass have been
performed in the \ttbar\ lepton plus jets channel using 162\,\invpb of
Tevatron Run~II data. The results are
$\mt=174.9^{+7.1}_{-7.7}(\mathrm{stat})\pm6.5(\mathrm{sys})\,\gevcc$
(TM),
$\mt=179.6^{+6.4}_{-6.3}(\mathrm{stat})\pm6.8(\mathrm{sys})\,\gevcc$
(MTM), and
$\mt=177.8^{+4.5}_{-5.0}(\mathrm{stat})\pm6.2(\mathrm{sys})\,\gevcc$
(DLM).\footnote{The results are consistent with an updated CDF
measurement in the dilepton channel reported after this conference:
$\mt=176.5^{+17.2}_{-16.0}(\mathrm{stat})\pm6.9(\mathrm{sys})\,\gevcc$.} 
The systematic error is dominated by jet energy uncertainties.  MTM
allows for an optimal compromise between statistical and systematic errors and
will be more powerful at higher accumulated luminosity.  The DLM result has
a considerably less statistical uncertainty than the other methods and
is competitive with the best CDF Run~I measurement~\cite{bib:cdfrun1}.

Tevatron is performing well and is expected to provide 400\,\invpb\ by
the end of 2004, 1\,\invfb\ by the end of 2005, and 4.4-8.5\,\invfb\
by the end of Run~II, thus leading to a significant improvement of the
statistical error.  A priority is the reduction of the jet energy
systematics.  The stated CDF goal for the total error is
$3\,\gevcc$. Substantial progress is expected soon, \eg\ due to recent
developments in the calorimeter simulation. Furthermore there is a substantial 
potential to refine the analyses.  Through these measurements, CDF has
established techniques that will provide a precise determination 
of the top quark mass in the future.

%%%%%%%%%%%%%%%%%%%%
%%% Bibliography %%%
%%%%%%%%%%%%%%%%%%%%


\small
\begin{thebibliography}{9} 
\bibitem{bib:topdiscovery} 
 F.~Abe \etal, \prl{74} (1995) 2626; S.~Abachi \etal, \prl{74} (1995) 2632.
\bibitem{bib:cdfrun1} T.~Affolder \etal, \prd{63} (2001) 032003.
\bibitem{bib:d0run1} 
B.~Abbott \etal, \prd{58} (1998) 052001;
S.~Abachi \etal, \prl{79} (1997) 1197.
\bibitem{bib:d0run1new} V.M.~Abazov \etal, \nat{429} (2004) 638.
\bibitem{bib:ewg} Tevatron EWG, hep-ex/0404010;
                  LEP EWG, \plet{{\rm B} 565} (2003) 61; 
updated Aug 2004: {\tt http://lepewwg.web.cern.ch/LEPEWWG/}.
\bibitem{bib:xsec} N.~Kidonakis and R.~Vogt, \prd{68} (2003) 114014.
\bibitem{bib:cdf2det} CDF Coll., FERMILAB-PUB-96-390-E (1996). 
\bibitem{bib:herwig} G.~Corcella \etal, JHEP {\bf 0101} (2001) 10.
\bibitem{bib:petra_merkel} CDF Coll., P.~Merkel, these proceedings.
\bibitem{bib:kde} D.~Scott, {\it Multivariate Density Estimation},
Wiley-Interscience, 1992.
\bibitem{bib:DLM} K.~Kondo \etal, J. Phys. Soc. Jap. {\bf 57} (1988) 4126;



\end{thebibliography}
\end{document}